\documentclass[aps,preprint,epsfig,rotate]{revtex4}
\usepackage{graphicx}
\usepackage{bm}
\usepackage{epsfig}

\begin{document}
\title{Asymptotic and interpolation series for the Coulomb three-body systems with unit charges}

\author{Alexei M. Frolov}
 \email[E--mail address: ]{alex1975frol@gmail.com} 


\affiliation{Department of Applied Mathematics \\
 University of Western Ontario, London, Ontario N6H 5B7, Canada}

\date{\today}

\begin{abstract}

Accurate mass-interpolation and mass-asymptotic formulas are derived for one- and two-center three-body ions with unit charges. The derived formulas 
are applied to predict accurate numerical values of the total energies of the ground (bound) $1^{1}S(L = 0)-$states in one-center atomic ions 
$X^{+} e^{-} e^{-}$ and analogous ground (bound) $1 s \sigma-$states in the two-center, quasi-adiabatic (or quasi-molecular) $X^{+} X^{+} e^{-}$ 
ions. We also discuss a few problems which currently remain unsolved for the $Q^{-1}$ expansions constructed for the ground (bound) states in 
few-electron atoms and ions.

\noindent 
PACS number(s): 36.10.-k and 36.10.Dr

\end{abstract}

\maketitle


\section{Introduction}

In this communication we investigate analytical dependence of the bound state properties of the Coulomb three-body systems with unit charges $X^{+} Y^{+} Z^{-}$ 
(or $X^{-} Y^{-} Z^{+}$) upon three particle masses $m_1 = m_X, m_2 = m_Y$ and $m_3 = m_Z$. As is well known the Hamiltonian of the non-relativistic Coulomb 
three-body system with unit electric charges ($X^{+} Y^{+} Z^{-}$) is written in the form
\begin{eqnarray}
 H = -\frac{\hbar^2}{2 m_e} \Bigl[ \frac{m_e}{m_1} \nabla^2_1 + \frac{m_e}{m_2} \nabla^2_2 + \frac{m_e}{m_3} \nabla^2_3 \Bigr] - \frac{e^2}{r_{32}} - 
 \frac{e^2}{r_{31}} + \frac{e^2}{r_{21}} \; \; \; , \; \; \label{Hamil3}
\end{eqnarray}
where $\hbar = \frac{h}{2 \pi}$ is the reduced Planck constant (or Dirac constant), $m_e$ is the electron mass, $m_1 = m_X, m_2 = m_Y$ and $m_3 = m_Z$ are the masses 
of the three point particles and $e$ is the electron's electric charge, while $m_e$ is the electron's mass. Here and everywhere below the notation $e$ stands for the 
electron. Formally, the Hamiltonian, Eq.(\ref{Hamil3}), coincides with the Hamiltonian of the charge-conjugate system $X^{-} Y^{-} Z^{+}$. This follows from the fact 
that the Hamiltonian $H$, Eq.(\ref{Hamil3}), is a quadratic function of the electric charge $e$. Rigorously, it can be shown by using the operator $\hat{Q}$ which 
changes signs of all electric charges to their alternative signs, i.e. $\hat{Q} e = - e$ and $\hat{Q} (Q e) = - e Q$. By applying this operator $\hat{Q}$ to the both 
sides of the Schr\"{o}dinger equation $H \Psi = E \Psi$ one can show that the energy spectra of bound states in the $X^{+} Y^{+} Z^{-}$ and $X^{-} Y^{-} Z^{+}$ systems 
are identical. In general, for Coulomb three-body systems with unit charges the operator $\hat{Q}$ is not conserved, but for some four-, six- and other similar Coulomb 
even-body systems with unit charges conservation of $\hat{Q}$ leads to an additonal physical symmetry in the system. For instance, for the three-body Ps$^{-}$ ion one 
finds $\hat{Q}$ (Ps$^{-}$) = Ps$^{+}$, which is a different physical system, while for the four-body Ps$_2$ quasi-molecule we always have $\hat{Q}$ (Ps$_2$) = Ps$_2$, 
and, therefore, this operator ($\hat{Q}$) represents an additional internal symmetry of the Ps$_2$ system. 

In this study we apply the atomic system of units, where $m_e = 1, e = 1$ and $\hbar = 1$. In these units the Hamiltonian, Eq.(\ref{Hamil3}), takes the following form   
\begin{eqnarray}
 H = -\frac12 \Bigl[ \frac{1}{m_1} \nabla^2_1 + \frac{1}{m_2} \nabla^2_2 + \frac{1}{m_3} \nabla^2_3 \Bigr] - \frac{1}{r_{32}} - \frac{1}{r_{31}} + 
 \frac{1}{r_{21}} \; \; \; , \; \; \label{Hamil33}
\end{eqnarray}
The formula, Eq.(\ref{Hamil33}), contains three physical parameters: $m_1, m_2, m_3$, which are the masses of three point particles expressed in the electron's mass $m_e$. 
Therefore, any wave function $\Psi$, which is the solution of the corresponding Schr\"{o}dinger equation $H \Psi = E \Psi$, is a `sufficiently smooth' (or analytical) 
function of these three parameters (or masses) $m_1, m_2, m_3$. This fact can rigorously be proven with the use of the well-known Poincare theorem about relation between 
singularities of the solution(s) of an arbitrary differential equation (e.g., the original Schr\"{o}dinger equation) with the singularities of coefficients in this equation 
(see, e.g., \cite{Mig1}). As follows from this fact the expectation values $\langle \Psi \mid A \mid \Psi \rangle$ computed with this wave function are also smooth functions 
of the three particle masses $m_1, m_2$ and $m_3$. Careful investigation of such functions and their mass-dependencies is the main goal of this study. Briefly, in this study 
we re-consider a number of important problems \cite{FroEr} which arise during investigation of analytical dependences of the bound state properties and actual wave functions 
of the Coulomb three-body systems with unit charges.

In general, any accurate and complete analsis of mass-dependencies of the bound state expectation values is a complex problem even for three-body Coulomb (quantum) systems 
with unit charges $X^{+} Y^{+} Z^{-}$, or $X^{-} Y^{-} Z^{+}$. Below, we restrict ourselves to the consideration of the two limiting cases of the three-body systems with 
unit charges which are most important for numerous applications: (a) one-center, quasi-atomic, two-electron (symmetric) systems $X^{+} e^{-} e^{-}$, where we have $m_1 = 
m_2(= m_e) \le m_3 = m_X = M$, and (b) two-center, adiabatic (symmetric) systems $X^{+} X^{+} e^{-}$ for which the following inequalities are obeyed $m_1 = m_2(= M) \gg 
m_3(= m_e)$. Note that the minimal particle mass in the both these cases equals $m_e$. The general theory of bound states in the Coulomb three-body systems with unit charges 
was developed more than 25 years ago in \cite{Posh}, \cite{FroBi} and \cite{FroBi1}. Some fundamental facts from that theory will be used in our analysis below. 

Our main goal in this study is to derive a number of simple, reliable and numerically stable interpolation and asymptotic formulas for the total energies of the ground states 
in one- and two-center three-body ions with unit charges. To derive such formulas in this study we shall apply numerical data from highly accurate computations of a large 
number of these ions. We also discuss a number of different situations which can be found for different Coulomb three-body systems, including the one-center atomic ions 
$X^{+} e^{-} e^{-}$ and two-center (but one-electron!) $X^{+} X^{+} e^{-}$ ions. In particular, we show that the $\frac{1}{M}$ series, which are used to represent actual 
computational data, should be constructed in the different forms for one- and two-center three-body ions with unit charges. Briefly, in this study we want to improve 
drastically our eralier results obtained for some of these atomic systems 30 years ago \cite{FroEr}.   

We also consider a few problems which currently exist for the $Q^{-1}$ expansions developed for the total energies of the two-, three- and four-electron atoms/ions. Such 
$Q^{-1}$-expansions were known (and used) in atomic physics, since the middle of 1930's when Hylleraas and later Bethe published their first papers about $Q^{-1}$-expansions 
for atomic systems (see, e.g., \cite{BS} and references therein). In our study we restrict ourselves to a few aspects of this old problem. Nevertheless, conclusions made 
below allows one to simplify future applications of the $Q^{-1}$ expansions for actual few-electron atomic systems and improve their overall accuracy.
 
\section{General Theory}

To investigate the mass-dependence of various expectation values $\langle \Psi \mid A \mid \Psi \rangle$ upon three particle masses one can directly use these three masses 
$m_1, m_2$ and $m_3$ as the actual parameters of the Coulomb three-body problem. However, in a large number of cases such a choice leads to a number of complications, since 
the area of mass variations is not restricted by any condition and each mass varies between 1 (or $m_e$) and $+ \infty$. This is not convenient in many applications, e.g., 
for graphical representation of the numerical results. As was found in a number of earlier studies it is more appropriate to apply a system of different parameters some of 
which vary between the two finite numerical values, i.e. they are the `compact' parameters. For instance, below for an arbitrary Coulomb three-body system we choose three 
parameters $\alpha, \beta$ and $\gamma$ which are defined as follows:
\begin{eqnarray}
 \alpha &=& \frac{2 m_1 m_2 m_3}{4 m_1 m_2 + m_3 (m_1 + m_2)} = \frac{2 m_1 m_2 m_3}{D} \; , \;  \beta = \frac{m_3 (m_1 + m_2)}{ 4 m_1 m_2 + m_3 (m_1 + m_2)} = \frac{m_3 
 (m_1 + m_2)}{D} \; , \; \nonumber \\
 \gamma &=& \frac{m_3 (m_1 - m_2)}{ 4 m_1 m_2 + m_3 (m_1 + m_2)} = \frac{m_3 (m_1 - m_2)}{D} \; , \; \label{mass}
\end{eqnarray}
where $D = 4 m_1 m_2 + m_3 (m_1 + m_2)$. The inverse relations take the form
\begin{eqnarray}
 m_1 = \frac{\alpha}{\beta - \gamma} \; \; \; , \; \; \; m_2 = \frac{\alpha}{\beta + \gamma} \; \; \; , \; \; \; m_3 = \frac{(\beta^{2} - \gamma^{2}) D}{\alpha} \; , \; 
 \label{massi}
\end{eqnarray}
Note that the two parameters $\beta$ and $\gamma$ are the dimensionless, compact parameters, while the first parameter $\alpha$ is a mass-dependent parameter which is 
proportional to the minimal particle mass. Without loss of generality, below we shall assume that such a minimal particle mass coincides with the electron mass $m_e$. The 
compact parameter $\gamma$ is often called the factor of assymetry (or parameter of assymetry). 

As mentioned above, in this study we restrict ourselves to the analysis of symmetric three-body systems, e.g., $X^{+} e^{-} e^{-}$ and $X^{+} X^{+} e^{-}$ ions and/or atoms. 
For such systems the parameter of assymetry ($\gamma$) equals zero identically. Furthermore, in this study our main goal is the investigation of the total energies $E$ of a 
number of the Coulomb three-body systems and their dependencies upon the masses of particles included in the system. The total energy of some bound state is the expectation 
value of the energy functional $E = \langle \Psi \mid H \mid \Psi \rangle$, where $\Psi$ is the unit norm wave function of this bound state and $H$ is the Hamiltonian of 
this system. In other words, the total energy $E$ is the expectation value of the Hamiltonian $H$ of the system computed with the bound state wave function $\Psi$ which is 
the solution of the non-relativistic Schr\"{o}dinger equation $H \Psi = E \Psi$. Investigation of other expectation values $\langle \Psi \mid \hat{A} \mid \Psi \rangle$ can 
be performed analogously. Here $\hat{A}$ is an arbitrary self-adjoint operator defined for this three-body system. In Quantum Mechanics the total energy of an arbitrary 
atomic (or Coulomb) system is always proportional to the electron rest mass, i.e. $E = a m_e$, where $a$ is a numerical constant. If the total energies of bound states are 
expresed in atomic units, then the coefficient $a$ equals $e^{4} \hbar^{-2}$. Now, we can write the following formula for the total energies of the one-center symmetric 
$X^{+} e^{-} e^{-}$ ions
\begin{equation}
  E = E(\alpha, \beta) = \alpha F(\beta) = \frac{m_e e^{4}}{\hbar^{2}} \Bigl(\frac{1}{1 + 2 y}\Bigr) F\Bigl(\frac{1}{1 + 2 y}\Bigr) \; \; \; \label{mass1}
\end{equation}
where $y = \frac{1}{M}$ and $F(x)$ is a regular (and real) function of the real argument $x$, while $M$ is the mass of the central quasi-nucleus $X^{+}$ expressed in the 
electron mass $m_e$, i.e. $m_X = M m_e = M$ (in atomic units). As follows from Eq.(\ref{mass1}) the total energy $E$ of any $X^{+} e^{-} e^{-}$ ion, expressed  in atomic 
units, realy depends upon one argument $\Bigl(\frac{1}{1 + 2 y}\Bigr)$ only. For the $X^{+} e^{-} e^{-}$ ions (from the Ps$^{-}$ ion up to the H$^{-}$ ion) this argument 
varies between 0 and $\frac13$. To derive the explicit form of the $F(x)$ function, Eq.(\ref{mass1}), we need to apply our numerical data obtained in highly accurate 
computations of actual three-body $X^{+} e^{-} e^{-}$ ions. This task is considered in the next Section.     

Analogous formula for the total energies $E$ of the symmetric two-center $X^{+} X^{+} e^{-}$ ions takes a slightly different form 
\begin{equation}
 E = E(\alpha, \beta) = \alpha f(\beta) =  \frac{m_e e^{4}}{\hbar^{2}} \Bigl(\frac{1}{2 + \frac{1}{M}}\Bigr) f\Bigl(\frac{1}{2 M + 1}\Bigr) = \frac{m_e e^{4}}{\hbar^{2}} 
 \Bigl(\frac{1}{2 + y}\Bigr) f\Bigl(\frac{y}{2 + y}\Bigr) \; \; \; \label{massA}
\end{equation}
where $y = \frac{1}{M} \ll 1$ and $f(x)$ is a different smooth and real function of its argument $x$. The explicit form of this function is determined in Section IV. For the 
two-center $X^{+} X^{+} e^{-}$ ion the parameter (or mass) $M$ coincides with the mass of one quasi-nuclei $X^{+}$, which is expressed in the electron mass $m_e$. The inverse
mass of the `heavy' particle $y = \frac{1}{M}$ is the small parameter which is used in our interpolation and asymptotic series, including Eq.(\ref{mass1}) and 
Eq.(\ref{massA}). However, despite some similarities between the one-center $X^{+} e^{-} e^{-}$ ions and two-center, adiabatic $X^{+} X^{+} e^{-}$ ions, there are a number of 
principal differences in the explicit form of the $E(\alpha, \beta)$ functions used to represent the total energies in each of these cases. These differences are explained in 
detail in the next two Sections. 

\section{One-center two-electron ions}  

In this Section we consider the two-electron $X^{+} e^{-} e^{-}$ ions. Note that all electrically charged particles in such ions have unit charges. Here we want to derive the 
explicit formula for the $E = \alpha f(\beta)$ function, Eq.(\ref{mass1}), which represents the total energies of the ground $1^{1}S_{e}-$states in these ions. In atomic 
units the formula for the $E = \alpha f(\beta)$ function, Eq.(\ref{mass1}), is written in the form 
\begin{eqnarray}
 E = E(\alpha, \beta) &=& \alpha F(\beta) = \Bigl(\frac{1}{1 + 2 y}\Bigr) F\Bigl(\frac{1}{1 + 2 y}\Bigr) = \frac{B_{1}}{1 + 2 y} + \frac{B_{2}}{(1 + 2 y)^{2}} + 
 \frac{B_{3}}{(1 + 2 y)^{3}} + \ldots \nonumber \\
 &=& \sum^{K}_{k=1} \frac{B_{k}}{(1 + 2 y)^{k}}  \; \; \; \label{massA1}
\end{eqnarray}
where $\alpha$ and $\beta$ are the continuous functions of $y$ (see above), $y = \frac{1}{M_X}$ and $B_{k}$ ($k$ = 1, 2, 3, $\ldots$) are the coefficients of this 
mass-interpolation series. Here we used the known fact that for the one-center $X^{+} e^{-} e^{-}$ ions the $f(\beta)$ function is a regular (or analytical) function of its 
argument $\beta$. In general, for the one-center and two-electron $X^{+} e^{-} e^{-}$ ions our interpolation series, Eq.(\ref{massA1}), provides very high numerical accuracy. 
This fact directly follows from the results of our calculations and very smooth form of the $E(y)$ dependence, Eqs.(\ref{mass1}) and (\ref{massA1}). By using highly accurate 
values of the total energies of the ground $1^{1}S-$states, determined for a large number of one-center $X^{+} e^{-} e^{-}$ ions (see Tables I and II), we can, in principle, 
obtain the correct numerical values of the first $K$ coefficients $B_k$ in the series, Eq.(\ref{massA1}). If such a number $K$ is relatively large, e.g., $K \ge 5$, then we 
can apply the derived formula, Eq.(\ref{massA1}), to predict accurate and highly accurate values of the total energies of the ground $1^{1}S-$states in the `new' $X^{+} e^{-} 
e^{-}$ ions without performing actual variational computations for these `new' two-electron ions. The overall accuracy of such predictions can be high (or even very high) and 
comparable to the best accuracy of actual computations. It is clear that the total number of the test systems $N_t$ must be larger (and even significantly larger) then the 
total number of determined coefficients $K$ in Eq.(\ref{massA1}). Therefore, in real applications one needs to use numerical codes based on the method of least squares, or 
other similar procedures. For instance, for the two-electron $X^{+} e^{-} e^{-}$ ions (see results in Tables I and II) we apply numerical data obtained for 25 different ions. 
This allows us to determine 15 - 21 first coefficients in the formula, Eq.(\ref{massA1}). In Table III we present only twenty (first) coefficients in Eq.(\ref{massA1}). 

As follows from Table III an obvious disadvantage of the interpolation formula, Eq.(\ref{massA1}), is the distribution of numerical values of the $B_i$ coefficients in this 
formula. Indeed, such a distribution is desribed by a function which has the bell-shape form (see Table III). Note also that the series, Eq.(\ref{massA1}), always contain an 
infinite number of terms even for $y = 0$ (the ${}^{\infty}$H$^{-}$ ion) and for $y = 1$ (the Ps$^{-}$ ion). This is not convenient for theoretical analysis in a number of 
actual cases. Therefore, it is better to develop a different new interoplation formula, which can also be used as the asymptotic formula. This can be made in a number of 
ways. One of these method, based on the results of our earlier approach \cite{FroEr}, is described here. In this approach the asymptotic-interpolation formula is written in 
the form
\begin{eqnarray}
 E = (1 - x) \sum^{K}_{k=0} {\cal B}_k x^{k} = x \Bigl[ E({}^{\infty}{\rm H}^{-}) + \sum^{K}_{k=1} {\cal B}_k z^{k} \Bigr] \; \; \; \label{massA5}
\end{eqnarray}
where $x = \frac{M_x}{M_x + 2 m_e} = \frac{M_x}{M_x + 2}, z = \frac{2 m_e}{2 m_e + M_X} = \frac{2}{2 + M_X}$ (in atomic units) and $E({}^{\infty}$H$^{-})$ is the total energy 
of the negatively charged ${}^{\infty}$H$^{-}$ ion (its numerical value is presented below). By using the numerical data for the $X^{+} e^{-} e^{-}$ ions mentioned above we 
determine the coefficients in the formula, Eq.(\ref{massA5}). Results of these computations (in $a.u$.) can be found in Table III. In Table III we present only eighteen 
(first) coefficients ${\cal B}_k$ in the interpolation (or asymptotic-interpolation) formula, Eq.(\ref{massA5}), by using numerical data for twenty five one-center, 
two-electron $X^{+} e^{-} e^{-}$ ions (see Tables I and II). In many applications the formula, Eq.(\ref{massA5}), is more convenient since for $M_X = \infty$ it contains only 
one term. Formally, the maximal accuracies for the both interpolation formulas, Eqs.(\ref{massA1}) and (\ref{massA5}), are very comparable to each other. For instance, for 
the total energy of the $X^{+} e^{-} e^{-}$ ion with $M_X = 60$ $m_e$ the predicted values are -0.51858484057893998101215 $a.u$. (for the both formulas), while the exact (or 
computational) value is $\approx$ -0.51858484057893998101310 $a.u$.

The interpolation formula, Eq.(\ref{massA1}), can be applied to represent the total energies $E$ of all three-body $X^{+} e^{-} e^{-}$ ions for $y$ is bounded between 1 and 
0 (or $M = \frac{1}{y}$ bounded between 1 and $\infty$). However, in some astrophysical studies, the total energies and other bound state properties of the hydrogen H$^{-}$ 
ions (isotopes) and systems close to them, e.g., the Mu$^{-}$ ion, are of great interest. For such systems we can replace the interpolation series, Eq.(\ref{massA1}), by the 
corresponsing asymptotic series (in atomic units)
\begin{eqnarray}
 E = b_{0} + \frac{b_{1}}{y} + \frac{b_{2}}{y^{2}} + \frac{b_{3}}{y^{3}} + \ldots = E({}^{\infty}{\rm H}^{-}) + \sum^{K}_{k=0} b_{k} y^{-k} \ldots \; \; \; \label{massA2}
\end{eqnarray}
which, in general, is more effective in applications to the H$^{-}$ ions and similar quasi-adiabatic systems. To derive the last formula we need to consider the term $t = 
\frac{1}{1 + 2 y}$ as the sum of geometrical progression, i.e. $t = \frac{1}{1 + 2 y} = 1 + \sum^{\infty}_{k=1} (-1)^{k} (2 y)^{k}$, since the $2 y$ value is very small (in 
fact, for all H$^{-}$ ions for the $2 y$ factor the following inequality: $2 y \le 1 \cdot 10^{-4}$ is always obeyed). Then, by using the expressions for the factors $t^{n}$ 
for $n = 1, 2, \ldots$ (written in terms of $y$) we can determine the coefficients in front of the positive powers of $y$. This can be achieved by reducing similar terms in 
the arising sums. The procedure leads to the new `asymptotic' series, Eq.(\ref{massA2}), for the negatively charged, hydrogen ions H$^{-}$ and close systems, 
Eq.(\ref{massA2}). These new asymptotic series are very effective in applications to the two-electron $X^{+} e^{-} e^{-}$ ions with very heavy $X^{+}$ nuclei (or 
quasi-nuclei). Note that the $b_{0}$ coefficient in the asymptotic formula, Eq.(\ref{massA2}), exactly conicides with the total energy of the negatively charged 
${}^{\infty}$H$^{-}$ ion, i.e. $b_0 = E({}^{\infty}$H$^{-})$ = -0.52775101 65443771 96590814 5667480(5) $a.u.$ The asymptotics series, Eq.(\ref{massA2}), explicitly written 
in terms of the inverse nuclear mass $\frac{1}{M}$ takes the form
\begin{eqnarray}
 E({}^{M}{\rm H}^{-}) = E({}^{\infty}{\rm H}^{-}) + \frac{b_{1}}{y} + \frac{b_{2}}{y^{2}} + \frac{b_{3}}{y^{3}} + \ldots = E({}^{\infty}{\rm H}^{-}) + \sum^{K}_{k=1} 
 b_{k} y^{-k} \ldots \; \; \; \label{massA3}
\end{eqnarray}     
To obtain the $b_{k}$ coefficients in this formula we need to use data from numerical computations of a number of actual H$^{-}$ ions. In reality, we have only three hydrogen 
ions with the finite (but very large!) nuclear masses: ${}^{1}$H$^{-}$ protium, ${}^{2}$H$^{-}$ (or D$^{-}$) deuterium and ${}^{3}$H$^{-}$ (or T$^{-}$) tritum. Highly accurate 
values of the ground state energies in these three ions are presented in Table II (all results are given in atomic units). In our numerical calculations of these ions we used 
the following `recent' nuclear/particle masses used in these computations:
\begin{eqnarray}
 m_e = 0.5109989461 \; \; \; , \; \; \; m_p = 938.2720813 \label{masses} \\
 m_{d} = 1875.612928 \; \; \; , \; \; \; m_t = 2808.921112 \nonumber
\end{eqnarray}
where all masses are expressed in the special high-energy mass units $MeV/c^{2}$. Note that these masses have been determined in recent high-energy experiments and they are 
currently recomended for scientific use by CODATA/NIST. 

In general, to determine the $b_{k}$ coefficients in Eq.(\ref{massA3}) to a good accuracy we need to use numerical data for 5 - 6 different hydrogen-like ions (at least). 
It is clear that in such calculations we cannot apply highly accurate results obtained for the same hydrogen isotopes (protium, deuterium and tritium) with slightly different 
(but very close!) masses. Indeed, the use of very close initial data lead to numerical instability of the whole solution process and sudden loss of the overall accuracy in all 
coefficients $b_1, b_2, b_3, \ldots$ of Eq.(\ref{massA3}). To avoid these troubles we have performed highly accurate computations of the ground $1^{1}S-$states in the three 
additional model ions ${}^{2000}$H$^{-}$, ${}^{4000}$H$^{-}$ and ${}^{6000}$H$^{-}$. Each of these systems is the two-electron, negatively charged hydrogen ion with one very 
heavy nucleus. The mass of such nucleus equals $M$ = 2000 $m_e$, $M$ = 4000 $m_e$ and $M$ = 6000 $m_e$, respectively. 

Results of highly accurate computations of the ground $1^{1}S-$states in these six one-center ions (in atomic units) can be found in Table II. Numerical data from Table II 
were used to determine the coefficients $b_{k}$ ($k$ = 1, 2, 3, $\ldots$) in the asymptotic formula, Eq.(\ref{massA3}). Highly accurate computations of the six hydrogen-like 
(two-electron) ions allow us to determine first four coefficients $b_k$ ($k$ = 1, 2, 3, 4) in Eq.(\ref{massA3}). These coefficients ($b_k$) can be found in the right column of 
Table III. By using these four coefficients and asymptotic formula, Eq.(\ref{massA3}), one can evaluate (to very high accuracy) the total energy of the ground $1^{1}S-$state 
of an arbitrary hydrogen-like ${}^{M}$H$^{-}$ ion with the nuclear mass $M$, where $M \ge 1500 m_e$. The accuracy of such an evaluation with the use of asymptotic fromula, 
Eq.(\ref{massA3}), is sufficient for all current and future experimental needs in modern astrophysics. For instance, by using the coefficients $b_k$ from this Table III one 
finds that the total energy of the ground $1^1S(L = 0)-$state in the ${}^{7000}$H$^{-}$ ions equals $\approx$ -0.5276709397607118862313(1) $a.u.$ which is very close to the 
actual value which can be evaluated from the following variational results $E({}^{7000}$H$^{-})(N = 4200) \approx$ -0.52767093976071188609100053 $a.u$. and 
$E({}^{7000}$H$^{-})(N = 4400) \approx$ -0.52767093976071188609105325 $a.u$. These numerical values can be compared to the total energies of such systems obtained in numerical
computations of other groups (see, e.g., \cite{Tel} and references therein).

\subsection{Computational Method}

In all highly accurate computations of the ground (bound) $1^{1}S-$states of the two-electron $X^{+} e^{-} e^{-}$ ions and $1 s \sigma-$states of the one-electron $X^{+} 
X^{+} e^{-}$ ions performed in this study we apply our `universal' exponential variational expansion \cite{Fro2001} which is written in one of the two following forms (see, 
e.g., \cite{Fro2001} and earlier references therein)
\begin{eqnarray}
 \Psi(r_{32}, r_{31}, r_{21}) = \sum_{i=1}^{N} C_{i} \exp(-\alpha_{i} r_{32} - \beta_{i} r_{31} - \gamma_{i} r_{21} - \imath \delta_{i} r_{32} - \imath e_{i} r_{31} - 
 \imath f_{i} r_{31})  \label{equ553} 
\end{eqnarray}
which is called the three-body exponential variational expansion in the relative coordinates $r_{32}, r_{31}, r_{21}$, or 
\begin{eqnarray}
 \Psi(u_1, u_2, u_3) = \sum_{i=1}^{N} C_{i} \exp(-\alpha_{i} u_1 - \beta_{i} u_2 - \gamma_{i} u_3 - \imath \delta_{i} u_1 - \imath 
 e_{i} u_2 - \imath f_{i} u_3)  \label{equ555} 
\end{eqnarray}
where $u_{1}, u_{2}, u_{3}$ are the three perimetric coordinates and all $6 N-$non-linear parameters $\alpha_{i}, \beta_{i}, \ldots, f_i$ ($i = 1, \ldots, N$) are real.
Optimization of these non-linear parameters and construction of the `short-term' cluster wave functions are described in detail in \cite{Fro2001} (see also earlier 
references therein). The perimetric coordinates have been introduced in physics of three-body systems by C.L. Pekeris in \cite{Pek1}. These three coordinates are simply 
(linearly) related to the relative coordinates
\begin{eqnarray}
  & & u_1 = \frac12 ( r_{21} + r_{31} - r_{32}) \; \; \; , \; \; \; r_{32} = u_2 + u_3 \nonumber \\
  & & u_2 = \frac12 ( r_{21} + r_{32} - r_{31}) \; \; \; , \; \; \; r_{31} = u_1 + u_3 \; \; \; \label{coord} \\
  & & u_3 = \frac12 ( r_{31} + r_{32} - r_{21}) \; \; \; , \; \; \; r_{21} = u_1 + u_2 \nonumber
\end{eqnarray}
where $r_{ij} = r_{ji}$. In contrast with the relative coordinates $r_{32}, r_{31}, r_{21}$ the three perimetric coordinates $u_1, u_2, u_3$ are truly independent of each 
other and each of them varies between 0 and $+\infty$. This drastically simplifies analytical and numerical computations of all three-body integrals which are needed for 
solution of the corresponding eigenvalue problem and for evaluation of a large number of bound state properties in an arbitrary three-body system. In actual applications 
the three last non-linear parameters (i.e. the $\delta_i, e_i, f_i$ parameters) in each of the basis function in Eq.(\ref{equ555}) can be chosen as arbitrary real numbers 
(each of them can be either positive, or negative, or zero), while the first three non-linear parameters $\alpha_{i}, \beta_{i}, \gamma_{i}$ must always be positive (real) 
numbers. The radial set of exponential basis functions must be a complete set. From here one finds a set of three additional conditions for the $\alpha_{i}, \beta_{i}, 
\gamma_{i}$ parameters. It can be shown that the set of exponential `radial' functions are complete, if (and only if) the three series of inverse powers of these parameters 
are divergent, i.e. the three following sums (or series): $S_1 = \sum_{i=1} \frac{1}{\alpha_{i}}, S_2 = \sum_{i=1} \frac{1}{\beta_{i}}$ and $S_3 = \sum_{i=1} 
\frac{1}{\gamma_{i}}$ are divergent when $i \rightarrow \infty$.

To perform highly accurate calculations of the ground $1 s \sigma-$state in the pure-adiabatic, two-center ${}^{\infty}$H$^{+}_{2}$ ion (see below) we applied our recently 
developed variational expansion in the prolate spheroidal coordinates (se, e.g., \cite{Mors} and \cite{LLELD}). Detail desciption of this expansion will be published 
elshewere. For the two-electron atomic systems discussed above (i.e. for the ${}^{\infty}$H$^{-}$ ion) and mentioned in the Appendix we applied the exponential variational 
expansion, Eq.(\ref{equ555}), in the perimetric coordinates, where all non-linear parameters $\delta_{i}, e_{i}$ and $f_{i}$ (for $i = 1, 2, \ldots, N$) were assumed to be 
equal zero, i.e., we deal with the `real' exponential expansion. For some of these two-electron ions (ions with $Q \le 10$) we could improve the overall accuracy of our 
numerical computations to stabilize up to 35 - 38 decimal digits (see Appendix). Unfortunately, our numerical code does cannot produce stable numerical results for the 
two-electron one-center ions with larger nuclear charges ($Q$) and for the Coulomb three-body systems where all particle masses are finite. Calculations of the three- and 
four-electron atomic systems and earlier numerical results obtained for such systems have been discussed in detail in \cite{Fro2016}.

\section{Adiabatic two-center ions with one bound electron} 

For the two-center, quasi-molecular $X^{+} X^{+} e^{-}$ ions we cannot simply repeat our procedure developed above, since for these ions there are a number of important 
differences with the one-center $X^{+} e^{-} e^{-}$ ions considered above. First of all, it is clear that each of the $X^{+} X^{+} e^{-}$ ions has two very heavy nuclei 
(or almost immovable `centers'), each of which is a positively charged, point particle. Briefly, we can say that in the limit $M_X \rightarrow \infty$ the `consequence' 
of the $X^{+} X^{+} e^{-}$ ions converges to the truly adiabatic ${}^{\infty}$H$^{+}_2$ ion which contains one bound electron. Therefore, the corresponding series 
constructed for the total energies and other expectation values are the asymptotic $\frac{1}{M}-$series, i.e. $M^{-1}$-series which are specifically designed to represent 
the total energies (and other bound state properties) of the two-center $X^{+} X^{+} e^{-}$ ions with large and very large nuclear masses $M_X$, i.e. $M_X \gg m_e = 1$. 
These series cannot be applied, in principle, to the Ps$^{-}$ ion and other similar three-body ions/systems. 

The second fundamental difference follows from the well-known fact that the truly adiabatic ${}^{\infty}$H$^{+}_2$ ion has an additional physical symmetry which leads to 
an additional degeneracy of the corresponding energy levels. Indeed, there are two additional (electronic) operators which commute with the Hamiltonian $H$ of the truly 
adiabatic ${}^{\infty}$H$^{+}_{2}$ ion. These operators are: (1) $z-$component of the total angular momenta $\hat{L}_z$ (or $\hat{L}^{2}_{z}$), and (2) the linear operator 
$\hat{\Lambda}$ which is a linear combination of the $\hat{L}^{2}$ operator and $z-$component of the Runge-Lentz operator $\hat{A}_{z}$, i.e. $\hat{\Lambda} = \hat{L}^{2} 
+ \sqrt{-2 E} R \hat{A}_z$, where $R$ is the inter-nuclear distance in the ${}^{\infty}$H$^{+}_2$ ion and $E$ is the total energy of the corresponding bound state. As 
mentioned above the operators $\hat{L}_z$ and $\hat{\Lambda}$ commute with the Hamiltonian of the truly adiabatic ${}^{\infty}$H$^{+}_{2}$ ion. However, the same operators 
do not commute with the Hamiltonians of the quasi-adiabatic ${}^{M}$H$^{+}_{2}$ ions. The lower physical symmetry of the actual ${}^{M}$H$^{+}_{2}$ ions (in comparison to 
the truly adiabatic ${}^{\infty}$H$^{+}_2$ ion) means splitting of the energy levels of the ${}^{\infty}$H$^{+}_2$ ion into $q$ different groups, where each group of 
eigenvalues corresponds to a different root of the eigenvalue equation which is, in fact, an algebraic (or polynomial-type) equation. Let us discuss this situation in 
detail. 

The actual Hamiltonians of the two-center $X^{+} e^{-} e^{-}$ ions are written in the following general from 
\begin{eqnarray} 
  H = - \frac12 \nabla^{2}_{e} - \frac{1}{2 M} \nabla^{2}_{1} - \frac{1}{2 M} \nabla^{2}_{2} + \frac{1}{R} - \frac{1}{r_{2e}} - \frac{1}{r_{1e}} = H_{0} - \frac{1}{2 M} 
 \Bigl(\nabla^{2}_{1} + \nabla^{2}_{2}\Bigr) = H_{0} + V \label{Hamt}
\end{eqnarray}
where $H_{0}$ is the Hamiltonian of the truly adiabatic ${}^{\infty}$H$^{+}_{2}$ ion, $R$ is the inter-nuclear distance in this ion and $V$ is the small perturbation which 
is a linear function upon the inverse nuclear mass $\tau = \frac{1}{M}$, where $\tau$ is the small parameter of this problem. Therefore, we can apply a simplified version 
of the complete theory developed in \cite{Kato} (see, e.g., Chapter II, \S 1, Section 7). In the linearized version of the complete theory \cite{Kato} all coefficients of 
the eigenvalue equation $Det( H(\tau) - E ) = 0$ are the polynomial function upon $\tau$. The eigenfunctions $E_{k}(\tau)$ correspond to different branches of the same 
algebraic function. If the eigenvalue equation $Det( H(\tau) - E ) = 0$ is irreducible, then all roots of this equation $E_{1}(\tau), E_{2}(\tau), \ldots, E_{N}(\tau)$ form 
one $N-$fold algebraic function and all eigenvalues belong to the same branch of one algebraic function. In such a case one finds no additional compications in comparison 
to the one-center $X^{+} e^{-} e^{-}$ ions discussed above. This means that one can still apply the asymptotic series which contain only integer powers of $\frac{1}{M}$ 
which were used, e.g., in Eqs.(\ref{massA1}) - (\ref{massA3}).

In the opposite case, i.e. when the eigenvalue equation $Det( H(\tau) - E ) = 0$ is reducible, all eigenvalues are splitted into a finite number of groups. Splitting of 
these eigenvalues into different groups is determined by the symmetries of the original (i.e. non-perturbed) problem ($\tau = 0$) and final problems (the cases when 
$\tau \ne 0$, or systems with perturbations). Since $\tau$ is a small parameter, then we deal with the additional symmetry in the unperturbed atomic system, i.e. in the 
truly adiabatic ${}^{\infty}$H$^{+}_{2}$ ion. In fact, in 99.99 \% of all applications the symmetry of the unperturbed atomic system (i.e. system with $\tau = 0$) is 
higher than the symmetry of analogous system with perturbation(s). When we move from the truly adiabatic ${}^{\infty}$H$^{+}_{2}$ ion to the `realistic' 
${}^{M}$H$^{+}_{2}$ ions (systems with lower symmetry), then it is easy to observe the actual splitting of all eigenvalues into a finite number of groups. As follows from 
here the regular power-series, e.g., Taylor and/or Maclaurin series, cannot be used to represent the function $f(x)$ from Eq.(\ref{massA}) for actual two-center, adiabatic 
$X^{+} X^{+} e^{-}$ ions. Indeed, for such atomic systems (ions) the higher symmetry of the original atomic system is broken and all algebraic roots of the eigenvalue 
equation $Det( H(\tau) - E ) = 0$ are splitted into a few different groups. Therefore, instead of the regular power series, one needs to apply the so-called Puiseux series 
of the $f(x)$ function which contains fractional powers of the argument $x$. In this case, Eq.(\ref{massA}), which now includes the Puiseux series, takes the form 
\begin{eqnarray}
 E(M) &=& \frac{M}{M + \frac12} \Bigl[ E(\infty) + \sum^{K_a}_{k=1} C_k \Bigl(\frac{1}{M}\Bigr)^{\frac{k}{q}} \Bigr] \; \; \label{Puis}
\end{eqnarray}
where $K_a$ is the maximal number of terms used in this series ($K_a$ can be finite, or infinite). The parameter $q$ is the Puiseux number (or Puiseux parameter), while 
the first term $C_0$ in the right-hand side of this formula coincides with the total energy of the ground (bound) $1 s \sigma-$state in the adiabatic 
${}^{\infty}$H$^{+}_{2}$ ion (non-perturbed system), i.e., $C_0 = E({}^{\infty}$H$^{+}_{2}) =  E(\infty) \approx$ -0.60263421 44949464 6150905(5) $a.u$. To apply the 
formula, Eq.(\ref{Puis}), to actual systems we need to determine the numerical value of the Puiseux number $q$. This Puiseux number $q$ can be found by using a few 
different approaches, e.g., we can apply the method based on the results of numerical computations of a large number of actual quasi-adiabatic ${}^{M}$H$^{+}_{2}$ ions. 
This method is called below the `numerical', or `experimental' approach. The second method is based on the fundamental result obtained early in the Born-Openheimer analysis 
\cite{BO} of the few- and many-electron systems which move in the field of the two heavy, positively charged atomic nuclei. An alternative approach (the `third' approach) 
is based on explicit computation of the exact `symmetry number' for the truly adiabatic H$^{+}_2$ ion. Then this `symmetry number' is used as the Puiseux number $q$ in 
actual computations based on the series Eq.(\ref{Puis}). Each of these three methods has its own advantages and disadvantages. The `numerical' method is discussed in detail 
in the next Section, while the additional symmetry of the ${}^{\infty}$H$^{+}_{2}$ ion will be considered elsewhere.  

First, by applying the method of Born-Oppenheimer \cite{BO} to the two-center, quasi-molecular $X^{+} X^{+} e^{-}$ ions one can show that the adiabatic parameter $\tau$ of 
the two-center approximation equals $\tau = \sqrt[4]{\frac{1}{M}}$. This means that the overall validity of the Born-Oppenheimer (or two-center) approximation vanishes 
(when the `nuclear' mass $M = M_X$ of particle $X$ decreases) for the $X^{+} X^{+} e^{-}$ ions as $\simeq \sqrt[4]{\frac{1}{M}}$. In other words, the total energies of the 
quasi-adiabatic ${}^{M}$H$^{+}_2$ ions are represented as the power series upon the small parameter $\tau = M^{-\frac14}$ and the Puiseux parameter $q$ equals four. Any 
other value(s) of $q$ will contradict the Born-Oppenheimer approximation. Finally, we obtain the following asymptotic formula for the total energies of the bound (ground) 
$1 s \sigma-$states in the one-electron quasi-adiabatic (or quasi-molecular) $X^{+} X^{+} e^{-}$ ions 
\begin{eqnarray}
  E(M) &=& \frac{M}{M + \frac12} \Bigl[ E(\infty) + \sum^{K_a}_{k=1} C_k \Bigl(\frac{1}{M}\Bigr)^{\frac{k}{4}} \Bigr] \; \; \label{sertwo} 
\end{eqnarray}
where $M$ is the mass of the model `proton' expressed in the electron mass $m_e$, $K_a$ is the total number of terms used to approximate the numerical data obtained in a 
series of $K$ highly accurate calculations of the different $X^{+} X^{+} e^{-}$ ions. Note that the derivation of the formula, Eq.(\ref{sertwo}), is essentially based on 
the Born-Oppenheimer approximation. Applications of the series, Eq.(\ref{sertwo}), with $q = 4$ to the adiabatic two-center ions and close systems are discussed in the next 
Section.   

\section{Computations and Conclusions}

In this Section we discuss numerical results of our computations obtained for a large number of bound states in the one-center $X^{+} e^{-} e^{-}$ ions and two-center, 
quasi-molecular $X^{+} X^{+} e^{-}$ ions. Below our main attention is devoted to the two-center, quasi-molecular $X^{+} X^{+} e^{-}$ ions, since the results of our numerical 
computations of the ground states in one-center $X^{+} e^{-} e^{-}$ ions have been discussed at the end of Section III. Note that all our calculations performed for the 
two-center, quasi-molecular $X^{+} X^{+} e^{-}$ ions are highly accurate (see Table IV). The overall accuracy of these our calculations can be evaluated as $\approx 1 \cdot 
10^{-21} - 1 \cdot 10^{-22}$ $a.u.$ For some three-body systems such an accuracy is even higher, while for other similar systems the overall accuracy of our calculations is 
slighlty lower. To construct the explicit mass-asymptotic formula, Eqs.(\ref{Puis}) - (\ref{sertwo}), we also apply results of our earlier computations of the two-center, 
quasi-molecular $X^{+} X^{+} e^{-}$ systems obtained in \cite{Fro2018}. 

By using the numerical data from our highly accurate computations one can determine a number of coefficients in the interpolation and asymptotic series derived above (see, e.g,
Eqs.(\ref{massA1}), (\ref{massA2}), (\ref{Puis}) and Eq.(\ref{sertwo}). In general, the more coefficients in such series can be determined, the better overall accuracy can be 
achieved for the interpolation and/or asymptotic series. However, in actual applications one always finds a number of restrictions. First, the absolute values of coefficients 
in such series must decrease and decrease rapidly. This means that the following inequalities for the coefficients in Eq.(\ref{sertwo}) must be obeyed: $\mid E(\infty) \mid > 
\mid C_2 \mid > \mid C_3 \mid > \ldots$. In reality, such coefficients always begin to increase after some value of $K = k_m$. Such a phenomenon is mainly related to the 
restricted accuracy of our numerical computations performed for the original systems (briefly, the accuracy of the original data). Indeed, the maximal numerical accuracy of 
prediction with the use of any interpolation and/or asymptotic formula cannot exceed the overall accuracy of the original data. In actual cases, such an accuracy of the 
asymptotic and/or interpolation series is always lower (and even substantially lower) than the accuracy of the original data. 	 

The situation when the coefficients in Eqs.(\ref{Puis}) - (\ref{sertwo}) begin to increase after some number $k_{m}$ is not rare and must be investigated carefully. Formally, 
if the $C_k$ coefficients increase slowly, then one needs to compare the $C_k M^{-\frac{k}{4}}$ terms in Eq.(\ref{sertwo}) by their absolute values. If the corresponding 
numerical values of the $\mid C_k \mid \cdot M^{-\frac{k}{4}}$ terms in Eq.(\ref{sertwo}) decrease when $k$ grows (it can be true, since the mass-parameter $2 M$ is very 
large), then we can still apply the series Eqs.(\ref{Puis}) - (\ref{sertwo}) to desribe the total energies of actual systems. However, if the absolute values of the $\mid C_k 
\mid M^{-\frac{k}{q}}$ coefficients rapidly increase (when the index $k$ grows), then we cannot apply the asymptotic series Eqs.(\ref{Puis}) - (\ref{sertwo}) to describe the 
total energies of actual systems. Sometimes the series, Eqs.(\ref{Puis}) - (\ref{sertwo}), can be used even in such cases, but results of such applications have a restricted 
physical sense. 

The arguments presented above were used to develop a different (numerical) mehtod which can be used to evaluate (or even determine) the actual Puiseux parameter for the 
quasi-adiabatic ${}^{M}$H$^{+}_{2}$ ions. In this method the parameter $q$ in the series, Eq.(\ref{Puis}) is varied until we obtain the regular asymptotic series, 
Eq.(\ref{sertwo}). Let us explain how this process works in reality. The variation process starts from $q = 1$ and continues (by increasing $q$ by unity) untill the series, 
Eq.(\ref{sertwo}), becomes the regular series which well describes all our computational data obtained for the two-center atomic ions. First, let us try the $q = 1$ value in 
the series, Eq.(\ref{Puis}). Results of numerical computations of the coefficients $C_k$ in the series, Eq.(\ref{Puis}), with $q = 1$ can be found in Table V. It is clear that 
the arising series is not a `regular' asymptotic series and its applications to actual systems can be describes as a disaster. Indeed, in this case the numerical values of the 
$C_k$ coefficients in the series, Eq.(\ref{Puis}), increase by a factor $\approx 1 \cdot 10^{4}$. Threfore, we cannot talk about any convergence of the asymptotic series, 
Eq.(\ref{Puis}), for $q = 1$. Analogous situations with the series, Eq.(\ref{Puis}), can be found for $q = 2$ and $q = 3$, but the absolute numerical values of the third, 
fourth, etc, coefficients in Eq.(\ref{Puis}) are smaller than we obsereved for $q = 1$ (see Table V). Briefly, one can say that in all these cases, i.e. for $q$ = 1 , 2 and 3, 
the series, Eq.(\ref{Puis}), do not converge et al.

However, for $q = 4$ the series, Eq.(\ref{Puis}), suddenly becomes the regular asymptotic series Eq.(\ref{sertwo}), which can be used to approximate all our computational data 
obtained for the two-center, quasi-adiabatic ${}^{M}$H$^{+}_{2}$ ions. Furthermore, in this case (i.e. for $q = 4$) the asymptotic series, Eq.(\ref{sertwo}), provides an accurate 
and numerically stable approximation for the original computational data (data used to determine the numerical values of the $C_k$ coefficients) and for all quasi-adiabatic 
${}^{M}$H$^{+}_{2}$ ions. For $q \ge 5$ the coefficients $C_k$ in the series, Eq.(\ref{Puis}), begin to increase to larger values (see Table V). For these values of $q$ 
the asymptotic series, Eq.(\ref{Puis}), do not converge, i.e. such series cannot be used to represent any actual computational data. Therefore, we can assume that the actual 
Puiseux parameter $q$ equals four in this case. It is interesting to note that the same conclusion is true, if the Puiseux parameter $q$ in Eq.(\ref{sertwo}) can be chosen as an 
arbitrary real number. Results of numerical computations of the $C_k$ coefficients in the series, Eq.(\ref{Puis}), for $q$ = 3.75, 3.90, 3.99, 4.10 and 4.25 can also be found in 
Table V. As folows from Table V the the coefficients $C_k$ in the arising series, Eq.(\ref{sertwo}), are minimal (by their absolute values) for $q = 4.00$, or the true numerical 
value of the Puiseux parameter $q$ equals four. In the previous Section based on the Born-Oppenheimer approximation we have found that the same value of the Puiseux parameter 
($q$ = 4) for the two-center, quasi-adiabatic ${}^{M}$H$^{+}_{2}$ ions. By using this Puiseux number, i.e. $q = 4$, we have determined a number of coefficients $C_k (k = 1, 2, 3, 
\ldots$) in the formula, Eq.(\ref{sertwo}). The Puiseux series with $q = 4$ can be used to predict the bound state energies of the `new' two-center, quasi-adiabatic 
${}^{M}$H$^{+}_{2}$ ions. By using our series, Eq.(\ref{sertwo}), we have evaluated the total energies $E$ of the two-center, quasi-adiabatic ${}^{M}$H$^{+}_{2}$ ions for a 
number of different (nuclear) masses $M$ which vary between $M$ = 50,000 $m_e$ and $M$ = 10,000,000 $m_e$. In fact, it is hard to imagine a point particle (or nucleus) with the
electric charge $+1$ and nuclear mass $M \ge$ 50,000 $m_e$. In other words, the problem of extrapolation of the total energies of the $1 s \sigma-$states in the 
${}^{M}$H$^{+}_{2}$ ions to the masses $M \ge$ 20,000 $m_e$ considered here is of a restricted and pure theoretical interest only.    

Numerical evaluations of the total energies of the ground (bound) $1 s \sigma-$states in the two-center, quasi-adiabatic (or quasi-molecular) $X^{+} X^{+} e^{-}$, where the 
`nuclear' masses $M$ vary between 50,0000 $m_e$ and 10,000,000 $m_e$ can be found in Table VI. As follows from the results of accurate numerical computations of the ground 
(bound) $1 s \sigma-$state in the ${}^{M}$H$^{+}_{2}$ ions with $M \ge$ 25,000 $m_e$ the overall accuracy of our predictions based on the formula, Eq.(\ref{sertwo}), is 
relatively low for $M \le$ 1,000,000 $m_e$. For larger `nuclear' masses the agreement between theoretical predictions and results of actual computations rapidly improving, 
but even for the ${}^{M}$H$^{+}_{2}$ ion the `nucleus' with $M$ = 10,000,000 the actual agreement can be observed only for the first 12 decimal digits. For our asymptotic 
formula, Eq.(\ref{sertwo}), which contains 1 + 8 = 9 terms (and only eight varied linear coefficients) such an agreement can be considered as good. However, it is clear that 
the accurate asymptotic formula, Eq.(\ref{sertwo}), must include a few dozens terms, e.g., forty and/or fifty terms. Briefly, the fraction $\frac{k}{4}$, which is the power of 
the last term in the formula, Eq.(\ref{sertwo}), must exceed eight (or even ten), i.e., $k \ge 32$ (or 40). If such a condition is obeyed, then we can used the formula, 
Eq.(\ref{sertwo}), for accurate numerical preditions of the total energies of the ground (bound) $1 s \sigma-$state in the ${}^{M}$H$^{+}_{2}$ ions with $M \ge$ 15,000 - 20,000 
$m_e$.     

Thus, we have developed a number of working asymptotic and interpolation series for the Coulomb three-body systems with unit charges. These formulas have been constructed for
the three-body $X^{+} Y^{+} Z^{-}$ systems (or ions) which are separated in the two classes: (a) all ions (or systems) bounded between the Ps$^{-}$ ion and ${}^{\infty}$H$^{-}$ 
ions, (b) the two-center, quasi-adiabatic (or quasi-molecular) $X^{+} X^{+} e^{-}$ ions close to the pure adiabatic ${}^{\infty}$H$^{+}_{2}$ ion. It is shown that the total 
energies $E$ for the ions of the first class (or (a)-class) can be approximated to a very good accuracy by the interpolation and asymptotic formulas, Eqs.(\ref{massA1}) and 
(\ref{massA5}). Both these series contains the integer powers of inverse nuclear masses. The corresponding asymptotic formulas for the second class (or (b)-class) include the
Puiseux series, Eqs.(\ref{Puis}) - (\ref{sertwo}). These series are very accurate in applications to the actual $X^{+} X^{+} e^{-}$ ions where $M_X \approx$ 5,000 - 10,000 
$m_e$. We also discovered the new criterion which can be used in actual applications to determine the actual numerical value of the Puiseux number $q$ in the asymptotic series. 
For the true Puiseux number all coefficients in the asymptotic series take their minimal (absolute) values possible. This simple creterion and numerical method based on this 
creterion can be very important in numerous applications, since the Puiseux series are often used to solve the system of differential equations with small perturbations, e.g., 
to determine the neutron's distribution(s) in various nuclear reactors, or predict trajectories of artificial satelites and other objects which move in the actual gravitational 
conditions. Unfortunately, the mentioned extremal properties of the truly Puiseux series were not known to me before this project started. \\    

{\bf Appendix}. \\

Let us briefly discuss the interpolation $Q^{-1}$-series for the Coulomb few-body atoms/ions, where $Q$ is the positive electric charge of the atomic nucleus $Q e$ (expressed 
in $e$). The total number of bound electrons in each of these ions is fixed and it is designated below as $N_e$. General theory of the $Q^{-1}$-series for atomic two-electron 
systems was created and extensively applied in the middle of the last century, e.g., in \cite{BS} and \cite{Eps} and references therein. Later the same problem was re-considered 
in a number of papers (see, e.g., \cite{Fro2016} and \cite{Shull} - \cite{Fro2010} and references therein). Our main interest in this study is related to the two-, three- and 
four-electron ions and neutral atoms. Here we want to discuss the problem of negatively charged ions which exist for some few-electron atoms and ions. Such stable, negatively 
charged ions, e.g., the H$^{-}$ and Li$^{-}$ ions, can be found in the two- and four-electron atomic systems. The both H$^{-}$ and Li$^{-}$ ions are stable in their ground 
singlet ${}^{1}S(L = 0)-$states. In our previous study \cite{Fro2016} we insisted that the results for the stable (ground) states in these negatively charged ions must be taken 
into account during construction of the numerical $Q^{-1}$-series for the two- and four-electron ions. Furthermore, it was suggested in \cite{Fro2016} that those $Q^{-1}$-series, 
which cannot reproduce accurate numerical results for the negatively charged ions to very good numerical accuracy, must be corrected in some way before they can be used in actual 
numerical evaluations. Since our paper \cite{Fro2016} was published we have found a number of arguments which support an alternative point of view. Therefore, it is important to 
re-consider the problem and re-evaluate the role of negatively charged ions which they play in the construction of accurate $Q^{-1}$-series for the few-electron ions.  

The problem discussed here is formulated in the following form. Let us consider the consequence of atoms/ions each of which contains $N_{e}$ bound electrons (this number does 
not change), while the electric charge of the central nucleus $Q$ (or $Q e$) is a variable. The Hamiltonian of such a non-relativistic atomic system is written in the form
\begin{equation}
 H = -\frac{\hbar^{2}}{2 m_{e}} \sum_{i=1}^{N_e} \nabla^{2}_{i} - Q e^2 \sum_{i=1}^{N_e} \frac{1}{r_{in}} + e^2 \sum^{N_e-1}_{j=1} 
 \sum^{N_e}_{i=2 (i > j)} \frac{1}{r_{ij}} \label{Ham}
\end{equation}
where $\nabla_{i} = \Bigl( \frac{\partial}{\partial x_{i}}, \frac{\partial}{\partial y_{i}}, \frac{\partial}{\partial z_{i}} \Bigr)$ and $i = 1, 2, \ldots, N_e$. In atomic 
units this atomic Hamiltonian takes the form 
\begin{equation}
 H = -\frac12 \sum_{i=1}^{N_e} \nabla^{2}_{i} - Q \sum_{i=1}^{N_e} \frac{1}{r_{in}} + \sum^{N_e-1}_{i=1} \sum^{N_e}_{i=2 (i > j)} \frac{1}{r_{ij}} \label{Ham1}
\end{equation}
Our goal below is to determine the total energies $E$ and non-relativistic wave functions $\Psi$ as the solutions of the Schr\"{o}dinger equation for the bound states $H \Psi 
= E \Psi$, where $H$ is the Hamiltonian of the atomic system, Eq.(\ref{Ham1}). The numerical parameter $E$ is a real, negative number (i.e. $E < 0$) and $\Psi$ is the unknown 
wave function. In general, it is clear that all negative eigenvalues $E_i$ and corresponding wave functions $\Psi_i$ of the Schr\"{o}dinger equation are the analytical 
functions of the nuclear charge $Q$. Based on this fact we can derive the following interpolation formula for the total energies of the ground ${}^{1}S(L = 0)-$states in a 
few-electron atoms/ions
\begin{eqnarray}
 E(Q, N_e) = a_2(N_e) Q^2 &+& a_1(N_e) Q + a_0(N_e) + b_1(N_e) Q^{-1} + b_2(N_e) Q^{-2} \nonumber \\
 &+& b_3(N_e) Q^{-3} + b_4(N_e) Q^{-4} + \ldots \; \; \; \label{Qexp2}
\end{eqnarray}
where all unknown coefficients are the analytical functions of the total number of bound electrons $N_e$ only, i.e. they do not depend upon $Q$. In this study we restrcit ourselves 
to the analysis of two-, three- ans four-electron atomic systems. Furthermore, we consider the $Q^{-1}$-expansions only for the total energies of ground $S(L = 0)-$states in these 
few-electron ions/atoms. This means that for two-electron ions we discuss the singlet $1^{1}S-$states, while for the three- and four-electron atoms and ions we deal with the doublet 
$2^{2}S-$ and singlet $2^{1}S-$states, respectively. Here we want to determine the unknown coefficients $a_2, a_1, a_0$ and $b_1, b_2, b_3, \ldots$ in the formula, Eq.(\ref{Qexp2}), 
by using the numerical results of highly accurate numerical calculations performed earlier (see, \cite{Fro2016} and references therein) for a large number of two-, three- and 
four-electron ions and atoms. Note that the series Eq.(\ref{Qexp2}) is a regular Loran series (in contrast with the series discussed in the previous Sections) which contains only 
integer powers of $Q$ and $Q^{-1}$. Definition of similar series for non-integer (but real!) values of $Q$ was also considered in the literature (see, e.g., \cite{Ivan}).

Originally, in 1930's the $Q^{-1}$-expansion was developed for the two-electron, positively charged ions and neutral atoms with the same number of bound electrons ($N_e$ = 2) and 
variable nuclear charge $Q$. Then almost twenty years later it was found that some negatively charged, atomic ions, and first of all, the H$^{-}$ ion(s), are also stable. The total 
energies of these ions have automatically been added to the basic data which were used to determine the coefficients in the $Q^{-1}$-expansion, Eq.(\ref{Qexp2}). Later, analogous 
negatively charged ions with the stable ground $S-$states were found in a number of iso-electron atomic systems, including the four-electron Li$^{-}$ ion. However, such negatively 
charged ions do not exist (as stable, bound systems) in the three- and five-electron atomic systems. The following question suddenly becomes interesting: why do we need to include 
the negatively charged ions in our series of calculations? Can highly accurate results for such ions increase the overall accuracy of our $Q^{-1}$-expansion for the two- and 
four-electron atomic systems. Otherwise, may be, such an inclusion will lead to an opposite result and we can loose the overall accuracy in the predicted total energies of the 
positively charged ions with large $Q$. In \cite{Fro2018} we have included the results obtained in highly accurate computations of the ground states of the H$^{-}$ and Li$^{-}$ ions 
in the set of standard `basis' two- and four-electron atomic systems. There a number of arguments which supprot this point of view. First of all, these negatively charged ions have 
the same internal electronic structure as other two- and four-electron ions/atoms, respectively. Another reason is more practical, since the $Q^{-1}$-series (constructed with the 
use of computational results obtained for positively charged ions and neutral atoms) provide relatively good approximations for the total energies of the corresponding negatively 
charged ions. The third argument is based on the fact that it is always better to include all known results for the two- and four-electron atomic systems (complete set of data), 
rather than exclude some of them. 

On the other hand, we have a number of alternative arguments and some of them are serious. In particular, each of these negatively charged ions have only one bound (ground) state, 
while all positively charged ions and neutral atoms (with the same number of bound electrons) have an infinite number of bound states each. In other words, the bound state spectra of 
each negatively charged ion is finite, but analogous spectra of the positively charged ions and neutral atoms is the Hilbert-Schmidt spectrum \cite{Fro2018A} which converges to the 
dissociation (or ionization) limit. Therefore, despite some similarity between electronic structure of these atomic systems, in reality we deal with the different iso-electron quantum 
systems. Another argument follows from earlier attempts to construct highly accurate interpolation formula Eq.(\ref{Qexp2}). It was found in some earlier studies that, if we do not 
include the total energies of the two- and four-electron negatively charged ions, then the arising interpolation formula, Eq.(\ref{Qexp2}), provides slightly better numerical accuracy 
for positively charged ions with larger nuclear charges $Q$. 

Finally, by analysing all these `pro-' and `contra-' arguments we arive to the conclusion that the explicit form of the $Q^{-1}$-formulas is determined by the goal of the original 
problem. For instance, if we want to approximate the total energies of bound states in heavy atoms and positively charge ions with large nuclear charges $Q$, then it is better to 
ignore all numerical data for the negatively charged ions which can be stable in this series of iso-electronic ions. However, if the aim of our numerical calculations is the 
derivation of the highly accurate $Q^{-1}$-expanion for light atoms and positively charged ions only, then it is obviously better to include in our computational data results obtained 
for the negatively charged ions (if they are stable). One actual example can be found in plasma physics, where there is a problem to describe the dense hydrogen-helim-lithium plasma 
at various temperatures (see, e.g., \cite{Fra}, \cite{Bru}, \cite{Fro2000} and references therein). In such problems one needs to introduce the `effective' nuclear charge $Q$ of these 
atomic mixtures, which is a fractional number bounded between 1 and 3. In such cases it is better to use the $Q^{-1}$-expansion, Eq.(\ref{Qexp2}), based on the results obtained, in 
part, for the negatively charged ions.      

In Table VII we present the numerical values of the first eleven coefficients $a_2, a_1, a_0$ and $b_1, b_2, b_3, \ldots$ in this formula, Eq.(\ref{Qexp2}), determined for the ground 
(bound) $S(L = 0)-$state in the two-, three- and four-electron ions and atoms. Such a number of terms (eleven) in the series, Eq.(\ref{Qexp2}), was chosen, since it provides the best 
numerical approximations for the known $a_2, a_1$ and $a_0$ coefficients in the series, Eq.(\ref{Qexp2}). Numerical values of these coeficients $a_2, a_1$ and $a_0$ have been predicted 
theoretically for $N_e$ = 2, 3 and 4 (in general, for arbitrary $N_e$ \cite{Fro2016}). Our calculations of the coefficients in the formula, Eq.(\ref{Qexp2}), have been performed with 
the use of the results obtained and discussed in \cite{Fro2016} (see, Table 1 in  \cite{Fro2016}). However, in this study the total energies of the first ten two-electron ions (from 
the hydrogen ${}^{\infty}$H$^{-}$ ion up to neon ${}^{\infty}$Ne$^{8+}$ ion) have been replaced with the more accurate numerical values. The overall accuraies of these `new' total 
energies vary between $\approx 2 \cdot 10^{-36}$ $a.u$. and $\approx 3 \cdot 10^{-38}$ $a.u$. Unfortunately, our computational method cannot produce relaible and stable results for the 
two-electron sodium ion and other similar two-electron ions with $Q \ge 11$. Note also, that currently, there is a general theory of the $Q^{-1}$-expansions developed for few-electron 
atoms and ions (see, e.g., \cite{Shull} - \cite{Fro2016} and references therein) which allows one to derive closed analytical expressions for the coefficients $b_k(N_e)$ in the series, 
Eq.(\ref{Qexp2}). In general, all these coefficients are determined as the functions of $Q$ and $N_e$, where $N_e$ is the total number of bound electrons. As mentioned in \cite{Fro2016}
in order to solve this problem accurately and completely we need to obtain more accurate data for three-, four- and five-electron atomic systems.

\newpage
\begin{table}[tbp]
   \caption{The total energies $E$ (in atomic units $a.u.$) of the ground $1^{1}S-$states of the two-electron $(X e e)^{-}$ ions. All masses of the hydrogen isotopes 
            have been taken from \cite{Fro2016A}.}
     \begin{center}
     \scalebox{0.85}{%
     \begin{tabular}{| l | l | l | l |}
       \hline\hline
 system  & $E$ & system & $E$ \\  
     \hline
 Ps$^{-}$${}^{(b)}$         & -0.26200507 02329801 07770400 325 & $(20m_e)^{+} e^{-} e^{-}$ & -0.50125455 80959544 801807  \\
 $(2m_e)^{+} e^{-} e^{-}$   & -0.34837166 58903586 377704       & $(30m_e)^{+} e^{-} e^{-}$ & -0.50975947 53251284 454137  \\
 $(3m_e)^{+} e^{-} e^{-}$   & -0.39214101 28604004 417462       & $(40m_e)^{+} e^{-} e^{-}$ & -0.51413087 24188312 365543  \\
 $(4m_e)^{+} e^{-} e^{-}$   & -0.41867330 89230921 649050       & $(50m_e)^{+} e^{-} e^{-}$ & -0.51679318 09776873 189919  \\
 $(5m_e)^{+} e^{-} e^{-}$   & -0.43649701 47071776 279064       & $(60m_e)^{+} e^{-} e^{-}$ & -0.51858484 05789399 810131  \\
        \hline\hline 
 $(6m_e)^{+} e^{-} e^{-}$   & -0.44930264 16011018 591330       & $(70m_e)^{+} e^{-} e^{-}$  & -0.51987293 00287915 022435  \\
 $(7m_e)^{+} e^{-} e^{-}$   & -0.45895089 40595627 468228       & $(80m_e)^{+} e^{-} e^{-}$  & -0.52084359 69480579 218423  \\       
 $(8m_e)^{+} e^{-} e^{-}$   & -0.46648274 15520219 800060       & $(90m_e)^{+} e^{-} e^{-}$  & -0.52160130 50180345 600732  \\
 $(9m_e)^{+} e^{-} e^{-}$   & -0.47252643 30300350 085221       & $(100m_e)^{+} e^{-} e^{-}$ & -0.52220921 01287829 565920  \\  
 $(10m_e)^{+} e^{-} e^{-}$  & -0.47748374 63598377 433885       & $(200m_e)^{+} e^{-} e^{-}$ & -0.52496409 08829953 751083  \\  
        \hline\hline 
 Mu$^{-}$                   & -0.52505480 65017307 7402573$^{(b)}$ & Mu$^{-}$                  & -0.52505480 65107558 9576070$^{(c)}$ \\  
 ${}^{1}$H$^{-}$            & -0.52744588 11197674 77071793        & ${}^{2}$H$^{-}$${}^{(d)}$ & -0.52759832 46897065 28596578 \\ 
 ${}^{3}$H$^{-}$${}^{(e)}$  & -0.52764904 82019207 33539418        & ${}^{\infty}$H$^{-}$      & -0.52775101 65443771 96590814 5667 \\ 
        \hline\hline 
  \end{tabular}}
  \end{center}
  ${}^{(a)}$Or $(m_e)^{+} e^{-} e^{-}$ in our current notations. \\
  ${}^{(b)}$The muon mass equals $M_{\mu}$ = (105.65836668/0.510998910) $m_e$ \\ 
  ${}^{(c)}$The muon mass equals $M_{\mu}$ = (105.6583745/0.5109989461) $m_e$ \\ 
  ${}^{(d)}$Also called the deuterium D$^{-}$ ion \\
  ${}^{(e)}$Also called the tritium T$^{-}$ ion \\
  \end{table}
\begin{table}[tbp]
   \caption{The total energies $E$ (in atomic units $a.u.$) of the ground $1^{1}S-$states of the two-electron hydrogen-like ions 
            ${}^{M}$H$^{-}$ where $M \gg m_e (= 1)$. For the protium, deuterium and tritium nuclei we used the `recent' set of nuclear masses, 
            Eq.(\ref{masses}).}  
     \begin{center}
     \scalebox{0.85}{%
     \begin{tabular}{| l | l | l |}
       \hline\hline
     hydrogen ion          & $E(N = 4200) $ & $E(N = 4400)$ \\  
      \hline
 ${}^{1}$H$^{-}$               & -0.52744588 11096971 99758650 & -0.52744588 11096971 99758699 \\  
        \hline
 ${}^{2}$H$^{-}$ (or D$^{-}$)  & -0.52759832 46845383 33600021 & -0.52759832 46845383 33600071 \\
        \hline
 ${}^{3}$H$^{-}$ (or T$^{-}$)  & -0.52764904 82021951 56625542 & -0.52764904 82021951 56625592 \\
        \hline\hline 
     model ion          & $E(N = 4200) $ & $E(N = 4400)$ \\
       \hline  
 ${}^{2000}$H$^{-}$ & -0.52747086 43674377 17613564 & -0.52747086 43674377 17613613 \\
       \hline\hline
 ${}^{4000}$H$^{-}$ & -0.52761089 96634719 11729427 & -0.52761089 96634719 11729476 \\
       \hline 
 ${}^{6000}$H$^{-}$ & -0.52765759 62215654 62168730 & -0.52765759 62215654 62168779 \\
        \hline\hline                     
  \end{tabular}}
  \end{center}
  \end{table}
\begin{table}[tbp]
   \caption{Numerical values of the $B_i$ and ${\cal B}_1$-coefficients from the interpolation formulas, Eqs.(\ref{massA1}) and Eqs.(\ref{massA5}), and 
            $b_i$-coefficients from the asymptotic formula Eq.(\ref{massA3}) (in atomic units).}  
     \begin{center}
     \scalebox{0.85}{%
     \begin{tabular}{| l | l | l | l | l |}
        \hline\hline
  $B_1$ =   -1.0410919060 & $B_6$   =   -3722.2073823724 & $B_{11}$ =  103830.6737884676 & $B_{16}$ = -13400.3074288148 & $b_1$ =  0.5606307984 \\
  $B_2$ =   -0.2621662929 & $B_7$   =   11351.7676116684 & $B_{12}$ = -107236.5623310799 & $B_{17}$ =   4148.1385021723 & $b_2$ = -0.6533093182 \\
  $B_3$ =   19.8524881732 & $B_8$   =  -27271.8028066598 & $B_{13}$ =   89845.6915073971 & $B_{18}$ =   -903.5006025786 & $b_3$ =  0.8411634494 \\
  $B_4$ = -170.6031166297 & $B_9$   =   52508.7030801499 & $B_{14}$ =  -60551.5208739613 & $B_{19}$ =    123.5257710913 & $b_4$ = -1.1956625006 \\
  $B_5$ =  934.4739314516 & $B_{10}$ = -81854.6083304213 & $B_{15}$ =   32357.0380042288 & $B_{20}$ =     -7.9763051000 &   ------------ \\
         \hline\hline
  ${\cal B}_1$ =  1.0000000000  &  ${\cal B}_5$ = -0.0407711935 &  ${\cal B}_9$    = -0.0086446997 & ${\cal B}_{13}$ =  0.0408891794 & ${\cal B}_{17}$ =  0.3586485140 \\
  ${\cal B}_2$ = -0.2474356173  &  ${\cal B}_6$ = -0.0244823770 &  ${\cal B}_{10}$ = -0.0073133662 & ${\cal B}_{14}$ = -0.1415203134 & ${\cal B}_{18}$ = -0.1496453527 \\
  ${\cal B}_3$ = -0.1304475478  &  ${\cal B}_7$ = -0.0158813569 &  ${\cal B}_{11}$ = -0.0041020681 & ${\cal B}_{15}$ =  0.2894077074 & -------------- \\
  ${\cal B}_4$ = -0.0716411979  &  ${\cal B}_8$ = -0.0112671393 &  ${\cal B}_{12}$ = -0.0160047510 & ${\cal B}_{16}$ = -0.4212134643 & -------------- \\
         \hline\hline
  \end{tabular}}
  \end{center} 
  \end{table}
\begin{table}[tbp]
   \caption{The total energies of the ground states (or $1 s \sigma-$states) of some two-center, quasi-molecular ions $X^{+}X^{+}e^{-}$ 
            in atomic units. The notation $N$ is the total number of basis functions used.}  
     \begin{center}
     \scalebox{0.85}{%
     \begin{tabular}{| l | l | l | l |}
      \hline\hline
 $N$  & $(M_p M_p e)^{+}$ ($M_p = 1000 m_e$) &  $(M_p M_p e)^{+}$ ($M_p = 1500 m_e$) & $(M_p M_p e)^{+}$ ($M_p = 2000 m_e$) \\  
     \hline
 4400 & -0.59509329 96958491 18445 & -0.59653116 02650119 64680 & -0.59737690 59509827 42559 \\
                  \hline\hline
 $N$  & $(M_p M_p e)^{+}$ ($M_p = 2500 m_e$) &  $(M_p M_p e)^{+}$ ($M_p = 5000 m_e$) & $(M_p M_p e)^{+}$ ($M_p = 20,000 m_e$) \\  
     \hline
 4400 & -0.59794915 10377439 18273 & -0.59935177 30213303 13969 & -0.60101160 89653286 42756 \\
         \hline \hline
  \end{tabular}}
  \end{center}
  \end{table}
\begin{table}[tbp]
   \caption{First eight coefficients $C_k (k = 1, 2, \ldots, 8)$ in the series, Eq.(\ref{sertwo}), as the functions of the Puiseux number $q$.}  
     \begin{center}
     \scalebox{0.65}{%
     \begin{tabular}{| l | l | l | l | l | l |}
      \hline\hline
       & $q = 1$ & $q = 2$ & $q = 3$ & $q = 4$ & $q = 5$ \\  
     \hline
 $C_1$ &  0.503555335826741904E+02  &  0.226679856454838830E+00  &  0.116777397542683385E-01  & -0.288279952111755252E-04  & -0.312944504617107894E-03  \\
 $C_2$ & -0.516976785662354465E+06  &  0.123480472684881327E+00  &  0.134584832246374006E+01  &  0.227777757120415354E+00  &  0.335481868440462636E-01  \\                
 $C_3$ &  0.356447543651520818E+10  & -0.680034128270731895E+01  & -0.216509396806670350E+02  & -0.147660324318790194E-01  &  0.528474939691323922E+00  \\
 $C_4$ & -0.144346782155324401E+14  &  0.521972440809636222E+03  &  0.345017655141785851E+03  &  0.223497022525278858E+00  & -0.154977877350846285E+01  \\  
            \hline
 $C_5$ &  0.344967136265602798E+17  & -0.251747425517562923E+05  & -0.385658588313645937E+04  & -0.981049840030698796E+00  &  0.498195298155850017E+01  \\
 $C_6$ & -0.476082978423350395E+20  &  0.736146916891474134E+06  &  0.277590579542561071E+05  &  0.377887807939899230E+01  & -0.110991430221682200E+02  \\                
 $C_7$ &  0.348481626646177275E+23  & -0.119504018677835612E+08  & -0.115160839984735794E+06  & -0.853801240668919370E+01  &  0.151175387430949113E+02  \\
 $C_8$ & -0.103925110513285117E+26  &  0.825676796728115634E+08  &  0.208916999290634665E+06  &  0.804015174950881455E+01  & -0.937480633559701394E+01  \\ 
         \hline\hline
       & $q = 3.75$ & $q = 3.9$ & $q = 3.99$ & $q = 4.1$ & $q = 4.25$ \\  
     \hline
 $C_1$ &  0.737994453846348230E-03  &  0.207414495242013203E-03  & -0.864438862593750137E-05  & -0.195104017315880558E-03  & -0.346763452465812072E-03 \\ 
 $C_2$ &  0.353590035472018275E+00  &  0.271679960362224221E+00  &  0.231836293459572370E+00  &  0.190787663885116703E+00  &  0.145842273621177605E+00 \\ 
 $C_3$ & -0.110601783373977927E+01  & -0.344969302076044258E+00  & -0.428369720568963930E-01  &  0.216483638625332325E+00  &  0.432387750905075345E+00 \\ 
 $C_4$ &  0.916523985602663618E+01  &  0.263710431994012443E+01  &  0.415414199917749233E+00  & -0.123542364078863511E+01  & -0.230876723950815602E+01 \\ 
           \hline
 $C_5$ & -0.563761417972587023E+02  & -0.145887873859087186E+02  & -0.200656649686163226E+01  &  0.634044988145992590E+01  &  0.107391512665118026E+02 \\ 
 $C_6$ &  0.230939442673164911E+03  &  0.549084402953823942E+02  &  0.744724460064279979E+01  & -0.209636767345732153E+02  & -0.330874660053183773E+02 \\ 
 $C_7$ & -0.553660592130517262E+03  & -0.121371293320691104E+03  & -0.162638811700379286E+02  &  0.408601253556753741E+02  &  0.603594349407760457E+02 \\ 
 $C_8$ &  0.585490313859065012E+03  &  0.118161798017468718E+03  &  0.152450302658413133E+02  & -0.357045842290086843E+02  & -0.492413779156570657E+02 \\ 
         \hline\hline
 \end{tabular}}
  \end{center}
  \end{table}
\begin{table}[tbp]
   \caption{The extrapolated (or predicted) total energies of the gound $1 s \sigma-$states in the symmetric two-center, quasi-molecular ions $X^{+}X^{+}e^{-}$ 
            as the functions of the `nuclear mass' $M_X = M_X m_e$ (in atomic units). The formula, Eq.(\ref{sertwo}), was used for our extrapolations.} 
     \begin{center}
     \scalebox{0.95}{%
     \begin{tabular}{| r | c | r | c |}
        \hline\hline
  $M_X$    &   $E$  &  $M_X$ & $E$ \\  
        \hline\hline        
   50,000  & -0.601612441245840117 &     750,000  & -0.602372097869040147 \\
  100,000  & -0.601913366940714779 &   1,000,000  & -0.602407318107939145 \\
  200,000  & -0.602125383675949501 &   2,000,000  & -0.602473944942273420 \\
  300,000  & -0.602219104988352996 &   5,000,000  & -0.602532997346062976 \\
  500,000  & -0.602312972176872087 &  10,000,000  & -0.602562729673433822 \\
      \hline\hline
  \end{tabular}}
  \end{center} 
  \end{table}
\begin{table}[tbp]
   \caption{Coefficients $a_2, a_1, a_0$ and $b_1, b_2, b_3, \ldots, b_7$ from Eq.(\ref{Qexp2}) determined for the two-, three- and four-electron
            atomic systems (in atomic units).}  
     \begin{center}
     \scalebox{0.85}{%
     \begin{tabular}{| l | l | l | l | l |}
        \hline\hline
         & $N_e = 2^{(a)}$  &  $N_e = 2^{(b)}$ & $N_e = 3$ & $N_e = 4$ \\  
        \hline\hline
  $a_2$  &  -1.0000000000  &  -1.0000000000  &  -1.1250000003  &  -1.2500000004 \\
  $a_1$  &   0.6250000003  &   0.6250000000  &   1.0228052530  &   1.5592742731 \\
  $a_0$  &  -0.1576664442  &  -0.1576664299  &  -0.4081684541  &  -0.8771232796 \\
           \hline
  $b_1$  &   0.0086993771  &   0.0086990424  &  -0.0164779988  &  -0.0422246650 \\
  $b_2$  &  -0.0008936865  &  -0.0008888931  &  -0.0419970616  &  -0.1834033937 \\
  $b_3$  &  -0.0009903348  &  -0.0010342904  &  -0.0316689195  &  -0.1530880415 \\
  $b_4$  &  -0.0008902046  &  -0.0006284997  &  -0.2088816524  &  -0.8893907819 \\
  $b_5$  &   0.0007051931  &  -0.0002949361  &   0.7294230783  &   3.1758933868 \\
  $b_6$  &  -0.0028392915  &  -0.0004898219  &  -3.0236928342  & -17.2676946335 \\
  $b_7$  &   0.0033699192  &   0.0003064144  &   5.9068491091  &  41.0756300675 \\
  $b_8$  &  -0.0022455447  &  -0.0005635117  &  -6.2448651281  & -54.9140616935 \\
         \hline \hline
  \end{tabular}}
  \end{center} 
 ${}^{(a)}$In this case the total energy of the ${}^{\infty}$H$^{-}$ ion is included in the set of basic data. \\ 
 ${}^{(b)}$In this case the total energy of the ${}^{\infty}$H$^{-}$ ion is excluded from the set of basic data. \\
  \end{table}
\end{document}